\documentclass[preprint,12pt]{elsarticle}

\usepackage{amssymb}
\usepackage{amsmath}

\usepackage{multirow}
\usepackage{subcaption}
\usepackage{algorithm}
\usepackage{algorithmic}
\usepackage{xurl}

\journal{Computer-Aided Design}

\begin{document}
	
	\begin{frontmatter}
		
		
		
		\title{Generating Sizing Fields for Mesh Generation via GCN-based Simplification of Adaptive Background Grids} 
		
		
		\author[1,2]{Xunyang Zhu}
		\ead{12424094@zju.edu.cn}
		
		\author[1,2]{Hongfei Ye\corref{cor1}}
		\ead{hfye@zju.edu.cn}
		
		\author[1,2]{Yifei Wang}
		\ead{Yifei_Wang@zju.edu.cn}
		
		\author[1,2]{Taoran Liu}
		\ead{taoranliu@zju.edu.cn}
		
		\author[1,2]{Jianjun Chen\corref{cor1}}
		\ead{chenjj@zju.edu.cn}
		
		\cortext[cor1]{Corresponding author}
		
		\affiliation[1]{organization={School of Aeronautics and Astronautics, Zhejiang University},
			addressline={38 Zheda Road}, 
			city={Hangzhou},
			postcode={310027}, 
			state={Zhejiang},
			country={China}}
		
		\affiliation[2]{organization={State Key Laboratory of Computer Aided Design and Computer Graphics, Zhejiang University},
			addressline={866 Yuhangtang Road}, 
			city={Hangzhou},
			postcode={310058}, 
			state={Zhejiang},
			country={China}}

		\begin{abstract}
			The sizing field defined on a triangular background grid is pivotal for controlling the quality and efficiency of unstructured mesh generation. However, creating an optimal background grid that is geometrically conforming, computationally lightweight, and free from artifacts like banding is a significant challenge. This paper introduces a novel, adaptive background grid simplification (ABGS) framework based on a Graph Convolutional Network (GCN). We reformulate the grid simplification task as an edge score regression problem and train a GCN model to efficiently predict optimal edge collapse candidates. The model is guided by a custom loss function that holistically considers both geometric fidelity and sizing field accuracy. This data-driven approach replaces a costly procedural evaluation, accelerating the simplification process. Experimental results demonstrate the effectiveness of our framework across diverse and complex engineering models. Compared to the initial dense grids, our simplified background grids achieve an element reduction of 74\%-94\%, leading to a 35\%-88\% decrease in sizing field query times. 
		\end{abstract}

		\begin{graphicalabstract}
			\includegraphics[width=\textwidth]{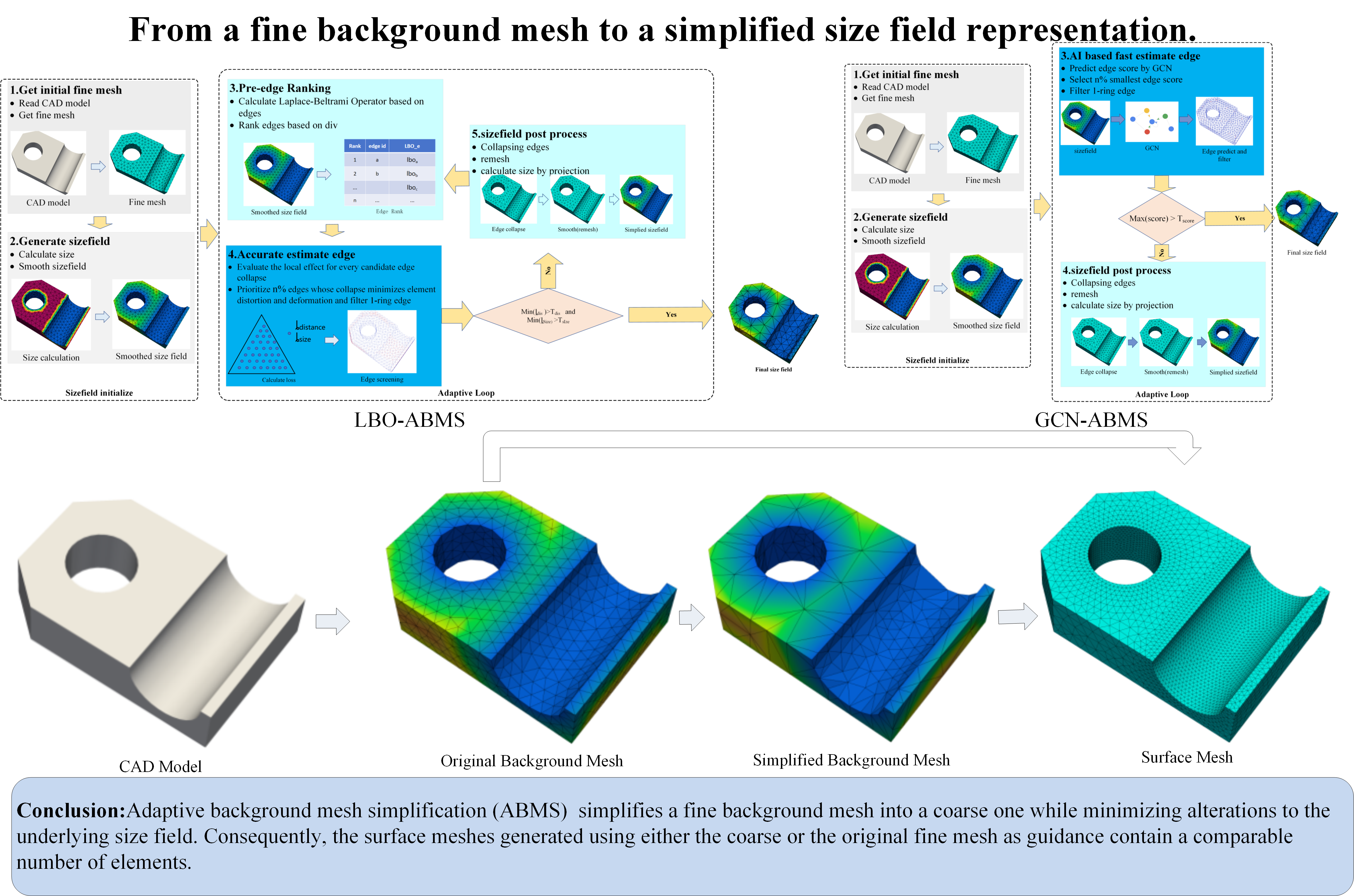}
		\end{graphicalabstract}
		
		\begin{highlights}
			\item A novel GCN-based framework for adaptive background grid simplification.
			\item Reformulation of grid simplification as a GCN-based edge classification task.
			\item Achieves 74-94\% element reduction in background grids while preserving quality.
			\item Accelerates sizing field queries by 35-88\% on complex engineering models.
		\end{highlights}
		
		\begin{keyword}
			Grid simplification \sep Sizing field \sep Background grid \sep Graph Convolutional Network \sep Unstructured mesh generation 
			
			
		\end{keyword}
		
	\end{frontmatter}
	
		
		\section{Introduction}
		For the various physical phenomena in nature that can be described by partial differential equations (PDEs)—such as fluid dynamics, solid mechanics, heat transfer, and electromagnetism—we have governing equations like the Euler and Navier-Stokes equations for fluid problems and Maxwell's equations for electromagnetic fields. In the face of complex geometries, strong nonlinearities, multiphysics coupling, and high-dimensional problems, purely analytical solutions are often intractable. By discretizing the continuous partial differential equations in both space and time (e.g., using methods such as the Finite Difference Method, Finite Element Method, Finite Volume Method, or Spectral Methods), we can simulate these physical laws on a computer. With the advancement of these numerical simulation techniques and the increasing complexity of engineering problems, the demands on the quality and density of the spatial discretization mesh are becoming progressively higher. A good mesh must simultaneously pursue two objectives: achieving the highest possible model accuracy with the fewest possible degrees of freedom (DoF). For a mesh with uniform element sizes, these two objectives are contradictory: reducing the element size improves model accuracy but inevitably increases the model's scale, and vice versa. The design of mesh density typically needs to consider numerous geometric and physical constraints that are independent of the meshing algorithm itself, such as narrow gaps, regions of high curvature, shock waves, boundary layers, chemical reaction fronts, and stress concentration zones. From the perspective of geometric accuracy, if we require the geometric error between the mesh model and the original model to be uniform across the domain, regions with high curvature need a finer mesh. Similarly, from the perspective of computational accuracy, if we require the computational error to be uniform, regions with high gradients (large variations in physical quantities) require a finer mesh. Therefore, using elements of varying sizes in different parts of the problem domain to adapt to changes in geometric curvature and physical gradients is an essential requirement for balancing the dual objectives of DoF and model accuracy. The resulting mesh is known as a variable-density mesh. In summary, we typically introduce a sizing field before mesh generation to exert more sophisticated control over the element sizes.
		
		There are two mainstream approaches for controlling the sizing field. One class of methods involves the user manually \cite{itoReliableIsotropicTetrahedral2004} or automatically specifying control sources based on physical or geometric constraints \cite{kaniaGeometricallyderivedBackgroundFunction2005}. These methods define a series of geometric sources (e.g., point sources, line sources, triangle sources \cite{aubryLinearSourcesMesh2013}), on which isotropic \cite{gridGeneration2008} or anisotropic \cite{aubryAnisotropicSourcesSurface2021} variations can be defined. The sizing field is then computed by blending the sizes from multiple sources. This approach often requires integration with a user interface and relies on manual operations.  
		
		The other class of methods is based on a background grid, where a sizing field is obtained by piecewise interpolation of size values defined on this grid. The size values are specified at the nodes or elements of the background grid. From another perspective, this can be seen as a source blending process, where the sizing field is a fusion of multiple size sources defined at each node of the background grid. The quality and density of the background grid are crucial parameters that determine the quality of the resulting sizing field. As the demand for mesh automation increases, background grid-based sizing field control has become mainstream due to its higher degree of automation, with some commercial software recently starting to support this approach (e.g., Pointwise). There are two main types of background grid methods. One is based on octrees \cite{deisterFullyAutomaticFast2004,quadrosComputationalFrameworkAutomating2010,pirzadehStructuredBackgroundgrids1993,quadrosComputationalFrameworkGenerating2005,quadrosSkeletonbasedComputationalMethod2004}. These methods construct an octree background grid of appropriate depth based on geometric information like curvature, storing size values at the octree nodes. A prolongation algorithm is then used to control the size field to satisfy size transition requirements. The other type is based on triangular grids \cite{xiaoAutomaticUnstructuredElementsizing2014,liuAutomaticSizingFunctions2021,lengParallelAutomaticMesh2024}. The general idea is that size values are defined at the nodes of a triangular grid, and the size at any other point is obtained through linear interpolation using barycentric coordinates. Typically, this method involves several steps: 1. The input triangular background grid is often an STL file generated by a geometry engine or a surface mesh generated by algorithms like the Advancing Front Technique (AFT). 2. Geometric and physical features are computed at the background grid nodes, and these are converted into size values using certain empirical formulas. 3. To satisfy size transition requirements, the size values on the background grid nodes need to be smoothed. A common practice is Laplacian smoothing, and for efficiency, some literature also employs iterative methods \cite{pippa2005gradh}. Finally, to query the size value at an arbitrary point (x,y,z) in space, the point is projected onto the background grid, and its size is calculated via barycentric interpolation. This process is usually accelerated using a spatial data structure such as an ADT, Octree, or k-d tree. 
		
		In recent years, the application of artificial intelligence (AI) in the field of mesh generation has become increasingly widespread, establishing AI-guided methods for unstructured mesh generation as a prominent research hotspot. 
		For instance, Freymuth et al.~\cite{freymuthIterativeSizingField2024} framed mesh generation as an imitation learning problem, where the goal is to mimic an expert-defined sizing field to guide the meshing process. 
		In a different approach, Xu et al.~\cite{xuImplicitGeometryNeural2025} employed a neural network to learn the mapping between the mesh sizing field and the signed distance field, thereby directing mesh creation. 
		Concurrently, AI-guided end-to-end mesh simplification has also emerged as a vibrant area of research, with significant contributions in recent years~\cite{potamiasNeuralMeshSimplification2022,potamias2022revisiting,xiangShapepreservingSimplificationMethod2022,wang2024simplified,wangHMSimNetHexahedralMesh2025,liuTetSimNetTetrahedralMesh2025}.
		
\begin{figure}[h]
			\centering
			\begin{subfigure}[b]{0.45\textwidth}
				\centering
				\includegraphics[width=0.75\textwidth]{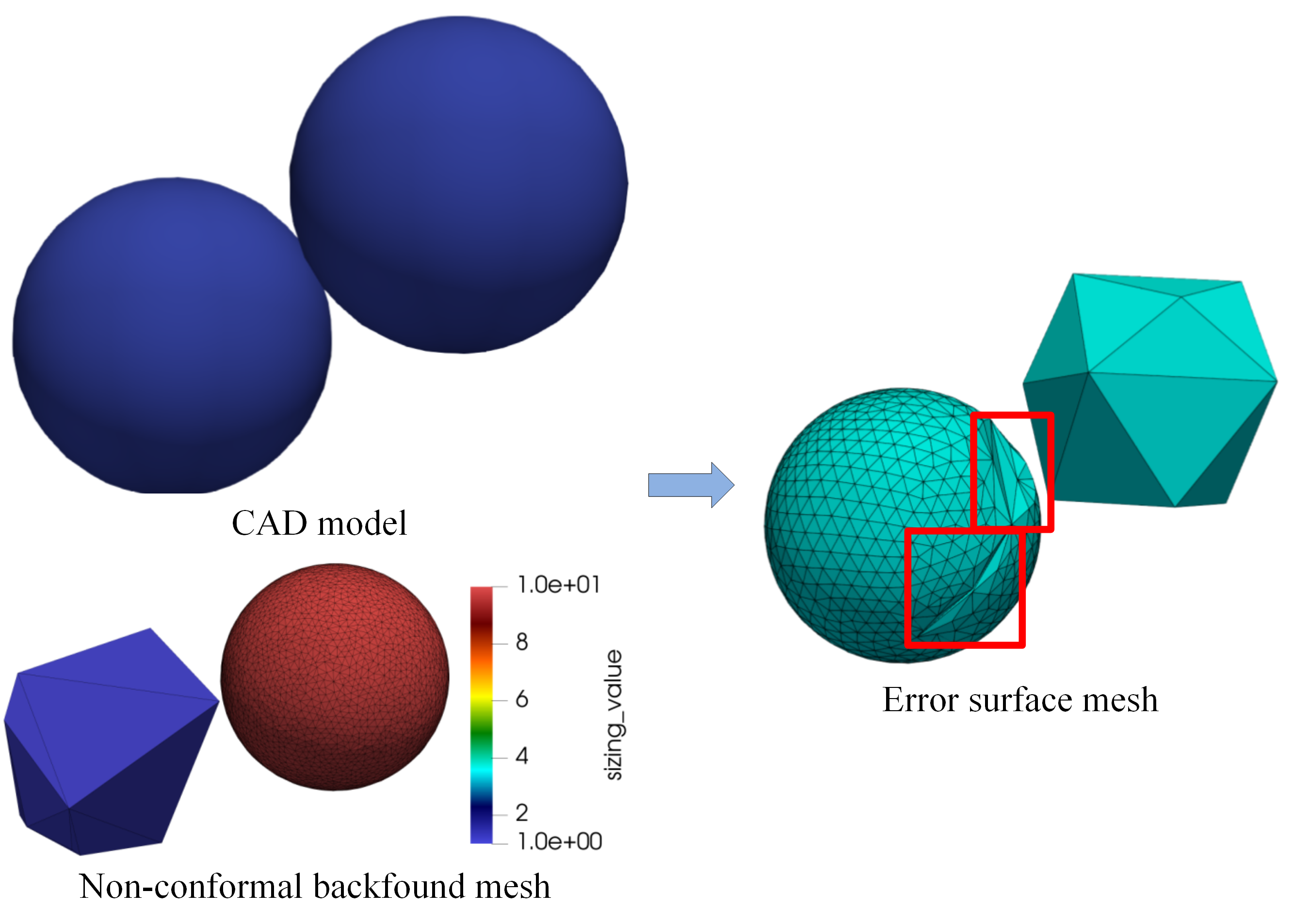}
				\caption{Non-conformal sizefield}
				\label{fig:non-conformal-bkgm}
			\end{subfigure}
			\hfill 
			\begin{subfigure}[b]{0.45\textwidth}
				\centering
				\includegraphics[width=\textwidth]{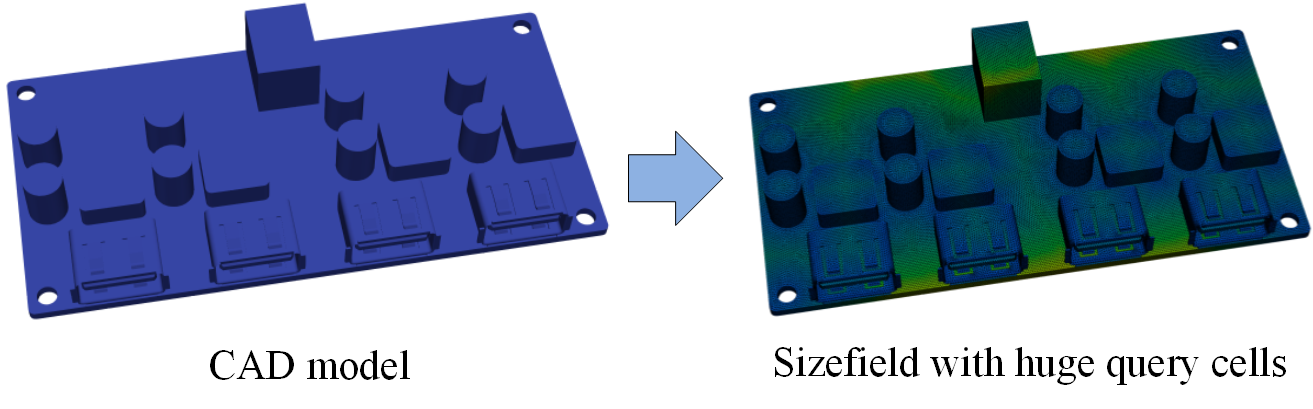}
				\caption{Size field with huge query cells}
				\label{fig:huge-cell}
			\end{subfigure}

			\vspace{1em} 
			
			\begin{subfigure}[b]{0.45\textwidth}
				\centering
				\includegraphics[width=0.75\textwidth]{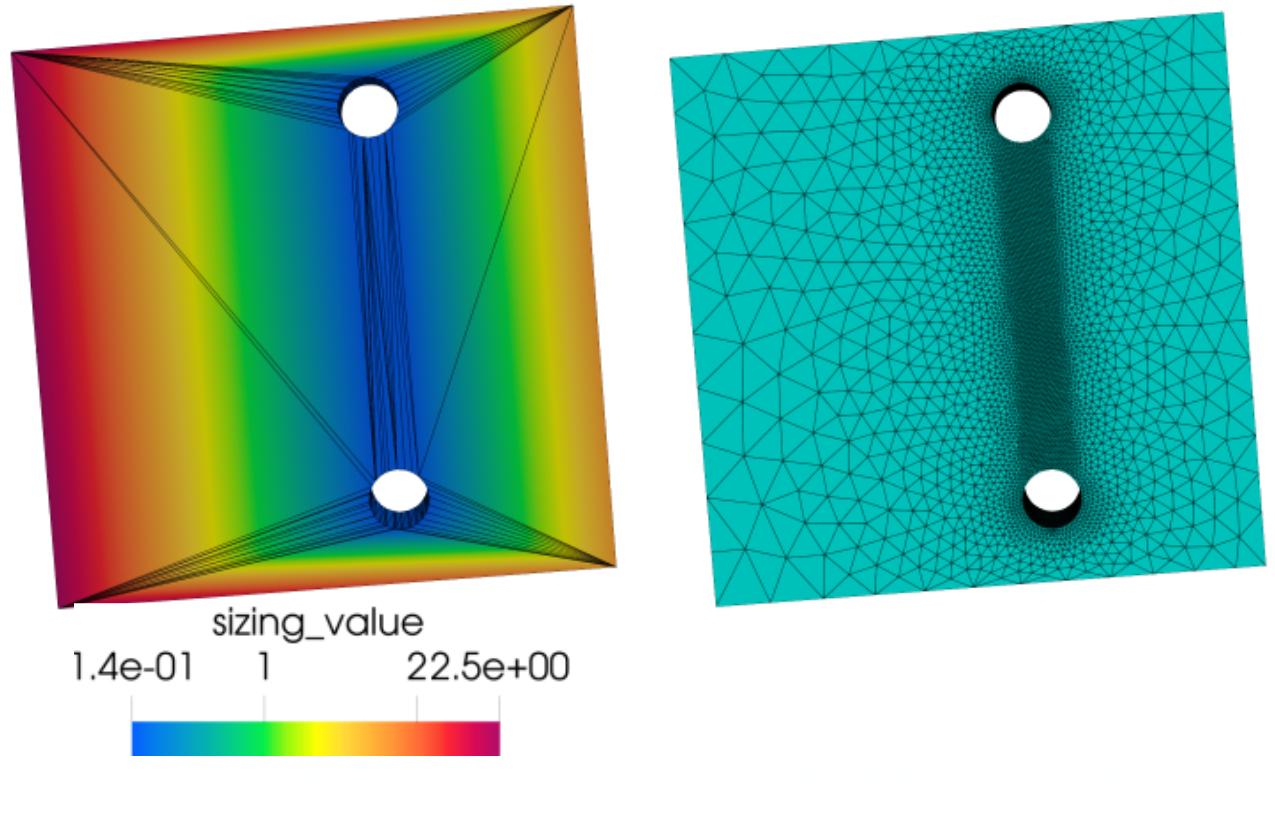}
				\caption{Banding sizefield}
				\label{fig:banding_condition}
			\end{subfigure}
			\hfill 
			\begin{subfigure}[b]{0.45\textwidth}
				\centering
				\includegraphics[width=0.75\textwidth]{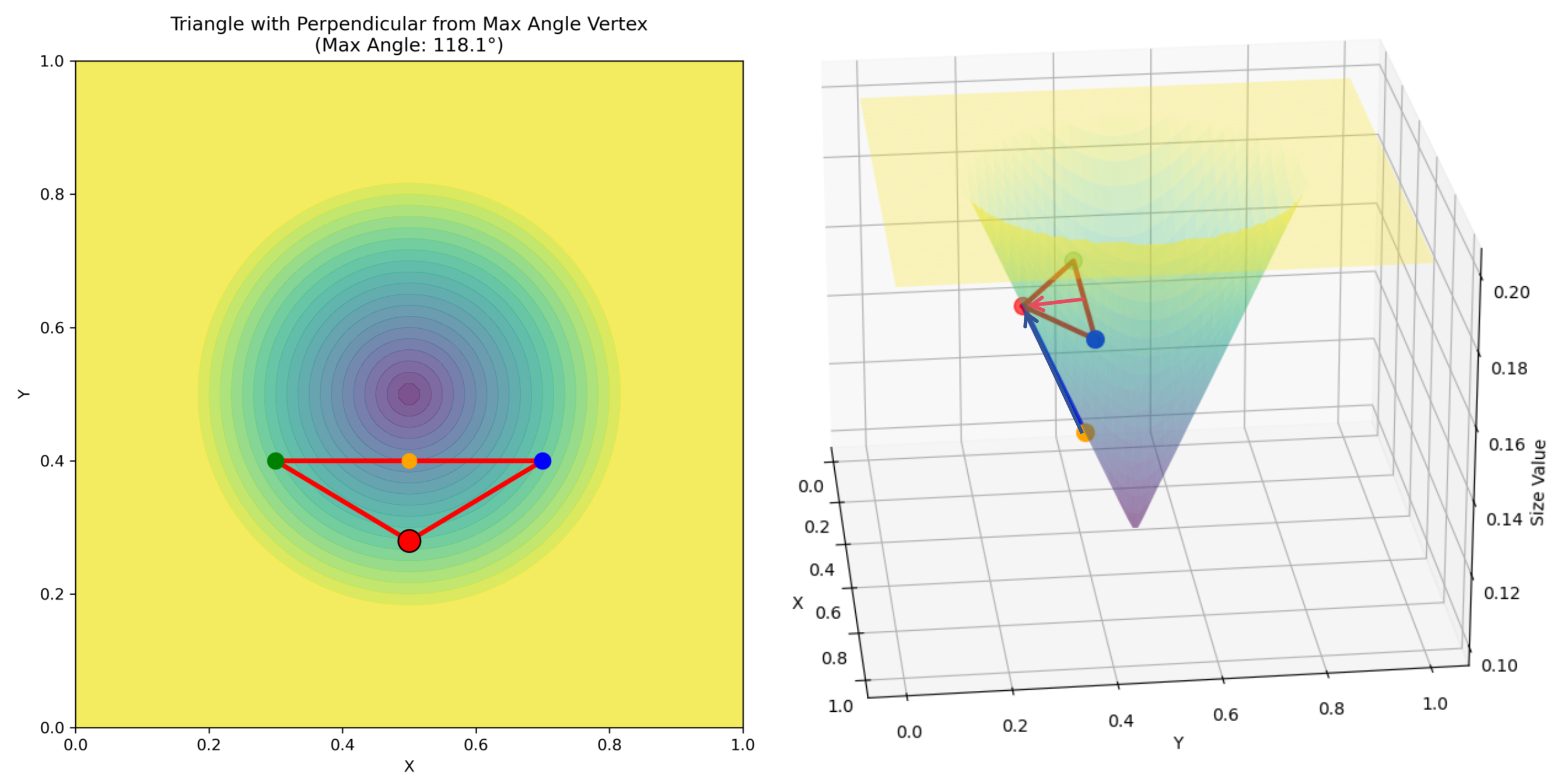}
				\caption{Sizefield with large angle query cells}
				\label{fig:large-angle}
			\end{subfigure}
			\caption{Several common issues with background grids.}
			\label{fig:bkgm-issue}
		\end{figure}

		A key question is: what constitutes a "good" triangular background grid? Per-Olof Persson \cite{perssonPdebasedGradientLimiting} framed the problem of sizing with octree/triangular background grids as solving for a gradient-limited size function. Assuming an ideal, continuous but non-differentiable size function $H^{*}$ that satisfies the gradient constraint, we need to find a function $H$ defined on the background grid that approximates $H^{*}$. In fact, this problem is very similar to the requirements for a mesh when solving PDEs. The following points are usually considered:
		
		a) The background grid needs to conform to the geometry to avoid geometric distortion caused by misprojection errors during the projection process. For example, as shown in Figure \ref{fig:non-conformal-bkgm}, a non-conformal background grid can cause size values from a near sphere to be incorrectly queried when generating the mesh on a nearby large sphere, resulting in poor mesh quality.
		
		b) Considering that the computational cost of high-quality sizing field smoothing methods \cite{perssonPdebasedGradientLimiting,pippa2005gradh,chenAutomaticSizingFunctions2017} often scales with the number of background grid elements, and that sizing field queries also depend on the background grid, the time and memory consumption are positively correlated with the number of elements (even with acceleration from a spatial tree). For instance, in the electromagnetic meshing scenario shown in Figure \ref{fig:huge-cell}, the mesh size can span up to four orders of magnitude. If the background grid for the sizing field is not optimized and a uniform grid is used instead, a vast number of background elements would be required to ensure a reasonable size transition. Therefore, it is necessary to control the number of background grid elements and avoid unnecessary refinement.
		
		c) To prevent the background grid from being too coarse to capture regions of rapid change in $H^{*}$, a phenomenon known as "banding," \cite{xiaoAutomaticUnstructuredElementsizing2014} the grid must have adequate density in regions where the gradient $\nabla H^{*}$ of $H^{*}$ changes sharply to ensure fidelity. As shown in Figure \ref{fig:banding_condition}, an inappropriate background grid can exhibit banding, which leads to an overly dense final mesh.
		
		d) The quality of the background grid is crucial for how well $H$ approximates $H^{*}$ \cite{shewchukWhatGoodLinear2002}. Unfortunately, if there are many large-angle elements, it is difficult to simultaneously reduce both $||H-H^{*}||$ and $||\nabla H^{*}-\nabla H||$. As illustrated in Figure \ref{fig:large-angle}, the yellow-green surface visualizes the gradient control field H*, while the red triangle highlights a single element within its discretization, H. When this element is an obtuse triangle, characterized by an excessively large maximum angle, its computed gradient can become zero. This result incorrectly suggests a zero gradient, failing to capture the true gradient of the underlying field. This implies that the quality of the background grid elements, especially those with large angles, should be optimized as much as possible.
		
	Considering the four aforementioned issues, methods based on triangular background grids typically use either STL files or surface meshes generated by algorithms such as AFT or Delaunay triangulation as the background grid. This approach leads to two significant problems. First, sizing field optimization on a background grid requires solving, or approximately solving, an optimization equation (e.g., via Laplacian smoothing). The typically poor quality of STL meshes can distort the solution of this equation, leading to abrupt jumps in the sizing field, a phenomenon known as "banding," as illustrated in Figure \ref{fig:banding_condition}. Moreover, the time required for subsequent sizing field queries is positively correlated with the number of mesh elements, which also degrades query efficiency. Traditional heuristic-based methods for background grid optimization, such as the work by Liu et al. \cite{liuAutomaticSizingFunctions2021}, identify banding edges by comparing the grid with a refined version and then performing remeshing. The edge selection process in these methods is not well-controlled. A more accurate approach is to iterate through all edges, remove an edge, re-apply the smoothing, and compare the results to determine if the simplification is acceptable. However, this brute-force approach for finding simplifiable edges is computationally expensive.
	
\begin{figure}[h]
		\centering
		\includegraphics[width=0.7\linewidth]{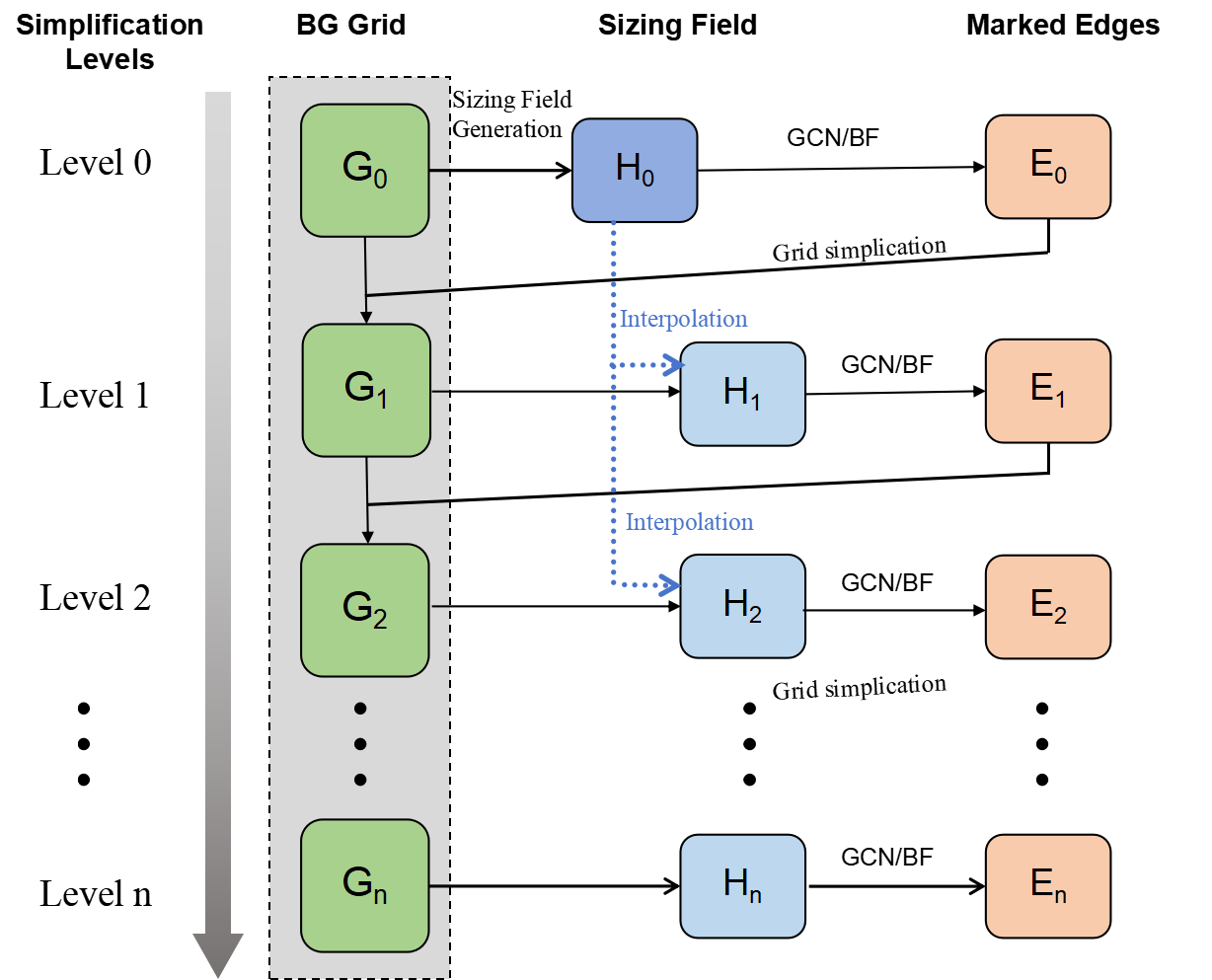}
		\caption{Schematic illustration of the adaptive background grid generation process.}
		\label{fig:adaptive-procedure}
	\end{figure}

	Addressing all the aforementioned issues simultaneously is a formidable and systemic challenge. To this end, this paper proposes a adaptive framework for simplifying triangular background grids as shown in Figure \ref{fig:adaptive-procedure}. The strategy begins with an initial, relatively refined background grid. By starting with a refined grid, we ensure that the requirements for geometric conformity (a), feature capturing (c), and element quality (d) are met as much as possible, allowing the subsequent process to focus on optimizing for criterion (b): minimizing the element count. Specifically, we use a GCN to predict edges that can be potentially removed and subsequently optimize the grid. Through this iterative process, a high-quality, simplified background grid is obtained.

	The primary contributions of this work are threefold:
	
	\begin{enumerate}
		\item We introduce a novel, GCN-based framework for adaptive background grid simplification. This framework generates geometry-conforming background grids that exhibit significantly fewer banding artifacts while using a smaller element count for queries. Compared to initial dense grids, our simplified grids achieve an element reduction of up to 94\%. When used for subsequent meshing, these simplified background grids guide the generation of surface meshes with over 50\% fewer elements than those derived from standard STL-based background grids. The method demonstrates strong generalization capabilities across diverse domains such as computational fluid dynamics and electromagnetics.
		
		\item Diverging from existing heuristic-based simplification methods, we reformulate the background grid simplification task as an edge classification problem within a GCN. To guide this process, we design a tailored loss function that holistically considers both geometric conformity and sizing field fidelity.
		
		\item We have compiled and will publicly release a benchmark dataset of background grids. This contribution is intended to foster further research, facilitate reproducibility, and promote technical advancements within the community.
	\end{enumerate}
	
	\section{Definition of 3D Manifold Sizing Field and Background Grid}
	Consider a three-dimensional Euclidean space $\Omega \subset \mathbb{R}^3$, within which a three-dimensional ideal scalar sizing field $H(\mathbf{x})$ is defined for $\mathbf{x} = (x, y, z) \in \Omega$. Concurrently, a geometric and physical size control field $H_0(\mathbf{x}')$ is defined on a three-dimensional submanifold $\Gamma \subset \Omega$, where $\mathbf{x}' = (x', y', z') \in \Gamma$, and this sizing field is defined exclusively on $\Gamma$. The values of the geometric and physical size control field $H_0(\mathbf{x}')$ are determined by local geometric features (such as curvature) and proximity to other features. Typically, smaller values are prescribed in regions of high curvature and close proximity to ensure that the mesh can capture these features. The definition of this field is often based on the meshing experience of simulation engineers. Based on this geometric and physical size control field, the computational model for the 3D field $H(\mathbf{x})$ can be defined as the steady-state solution of a Hamilton-Jacobi equation \cite{perssonPdebasedGradientLimiting}, given by:
	\begin{multline}
		H(\mathbf{x}) = \lim_{t \to \infty}	H_t(\mathbf{x}) =  \min_{\mathbf{x}' \in \Gamma} \left[ H_0(\mathbf{x}') + \beta \cdot d_{\Gamma}(\mathbf{x}, \mathbf{x}') \right], \\ \quad \beta \in [1, +\infty)
		\label{eq:conti_definition}
	\end{multline}
	where $d_{\Gamma}(\mathbf{x}, \mathbf{x}')$ denotes the geodesic distance from point $\mathbf{x}$ to point $\mathbf{x}'$ on the manifold $\Gamma$, and $\beta$ is a transition rate parameter that controls the range of influence of the geometric and physical size control field. This model exhibits the following properties:
	\begin{enumerate}
		\item $H_0(\mathbf{x}')$ represents local size influence: When a point $\mathbf{x}$ is close to the submanifold $\Gamma$ where the size control field is defined, $H(\mathbf{x})$ is primarily influenced by the value of $H_0(\mathbf{x}')$ at the nearest point on $\Gamma$, as the geodesic distance $d_{\Gamma}(\mathbf{x}, \mathbf{x}')$ is minimized.
		\item $d_{\Gamma}(\mathbf{x}, \mathbf{x}')$ represents the global smoothing requirement: When a point $\mathbf{x}$ is far from the submanifold $\Gamma$, smaller size values $H_0(\mathbf{x}')$ at distant locations on $\Gamma$ can still influence $H(\mathbf{x})$ through the minimization process, despite the large geodesic distance $d_{\Gamma}(\mathbf{x}, \mathbf{x}')$. This ensures a smooth transition across the entire field.
		\item The transition rate parameter $\beta$ controls the smoothness of the size transition. A larger value of $\beta$ causes the influence of distant $H_0(\mathbf{x}')$ values on $H(\mathbf{x})$ to decay more rapidly.
	\end{enumerate}
	However, maintaining the entire continuous field on $\Gamma$ is computationally challenging. In practice, a discrete size control strategy is typically employed. Specifically, a triangular background grid $M=(V,E,F)$ is constructed, where $V$ is the set of vertices, $E$ is the set of edges, and $F$ is the set of faces. A discrete sizing field is defined on this background grid, where each vertex $v\in V$ is assigned a size value, i.e., $H_M:V\rightarrow\mathbb{R}$. For any point $\mathbf{x}$ within a face of the background grid $M$, its size value is obtained through barycentric interpolation:
	\begin{equation}
		H_M(\mathbf{x})=\sum_{i=1}^3\lambda_i(\mathbf{x})\cdot H_M(v_i)
		\label{eq:discr_definition}
	\end{equation}
	where $\{v_1,v_2,v_3\}$ are the three vertices of the triangular face containing the point $\mathbf{x}$, and $\lambda_i(\mathbf{x})$ is the barycentric coordinate of $\mathbf{x}$ with respect to vertex $v_i$, satisfying $\sum_{i=1}^3\lambda_i(\mathbf{x})=1$ and $\lambda_i(\mathbf{x})\geq0$.
	
	This background grid-based approach discretizes the continuous field from $\Gamma$ onto the grid vertices, which preserves the primary characteristics of the original model while significantly reducing computational complexity. The topological structure $(V,E,F)$ of the background grid provides an efficient spatial indexing mechanism, allowing for the rapid localization of the triangular face containing any point $\mathbf{x} \in \Omega$ and the subsequent calculation of its size value via barycentric interpolation. The objective of the background grid-based method is to achieve rapid generation and querying of the sizing field while minimizing the loss of accuracy, i.e., ensuring that the deviation between the discrete sizing field $H_M(\mathbf{x})$ and the continuous field $H(\mathbf{x})$ is small. The solution procedure is divided into two steps:
	\begin{enumerate}
		\item \textbf{Initial Sizing Field Generation:} Based on the geometric and physical size control field 
		$H_0(\mathbf{x}')$ on the submanifold $\Gamma$, initial size values $H_{M,0}(v) (v \in V)$ are obtained by sampling at the vertices $V$ of the background grid.
		\item \textbf{Sizing Field Smoothing and Optimization:} The discrete sizing field is smoothed under the constraint that the vertex size values are non-increasing (i.e., for \(\forall v \in V\), the optimized size value is not greater than the initial sampled value). Common methods for this step include iterative relaxation \cite{liuAutomaticSizingFunctions2021} and the global equation solving approach proposed by our team \cite{xiaoAutomaticUnstructuredElementsizing2014}. Notably, our team has proven that when a gradient constraint of \(\beta\) is imposed on the triangular faces $F$, this sizing field optimization problem is equivalent to a convex optimization problem with a strictly convex objective function, which guarantees the uniqueness of the optimal solution.
	\end{enumerate}
	
	However, using $H_M$ to approximate $H$ introduces challenges similar to those in traditional finite element meshing: if the elements of $M$ are too large in critical regions (e.g., where the gradient of $H$ is large), the resulting sizing field $H_M$ can be distorted; if $M$ does not conform closely to $\Gamma$, geometric inaccuracies can occur; and if the number of faces $|F|$ is too large, the process becomes computationally expensive. A high-quality background grid $M$ should be geometry-conforming, have a low element count, and exhibit smooth and reasonable size transitions. Due to the lack of prior knowledge, selecting a background grid that meets all these criteria is difficult.
	
	To address these issues, we introduce a GCN-based optimization framework for $M$, which can automatically learn the characteristics of the sizing field based on geometric features and other information.
	
	\section{Methodology}
	This section details our approach to adaptive background grid coarsening. We first describe a procedural framework that relies on a pre-ranking algorithm to identify edges for potential collapse (Section \ref{sec:procedural}). Building upon this, we then present a more advanced, data-driven framework that leverages a GCN to learn optimal coarsening policies directly from grid data (Section \ref{sec:gnn}).
	
	\subsection{Procedural Framework for Adaptive Simplification}\label{sec:procedural}
	\begin{figure*}[h]
		\centering
		\includegraphics[width=\textwidth]{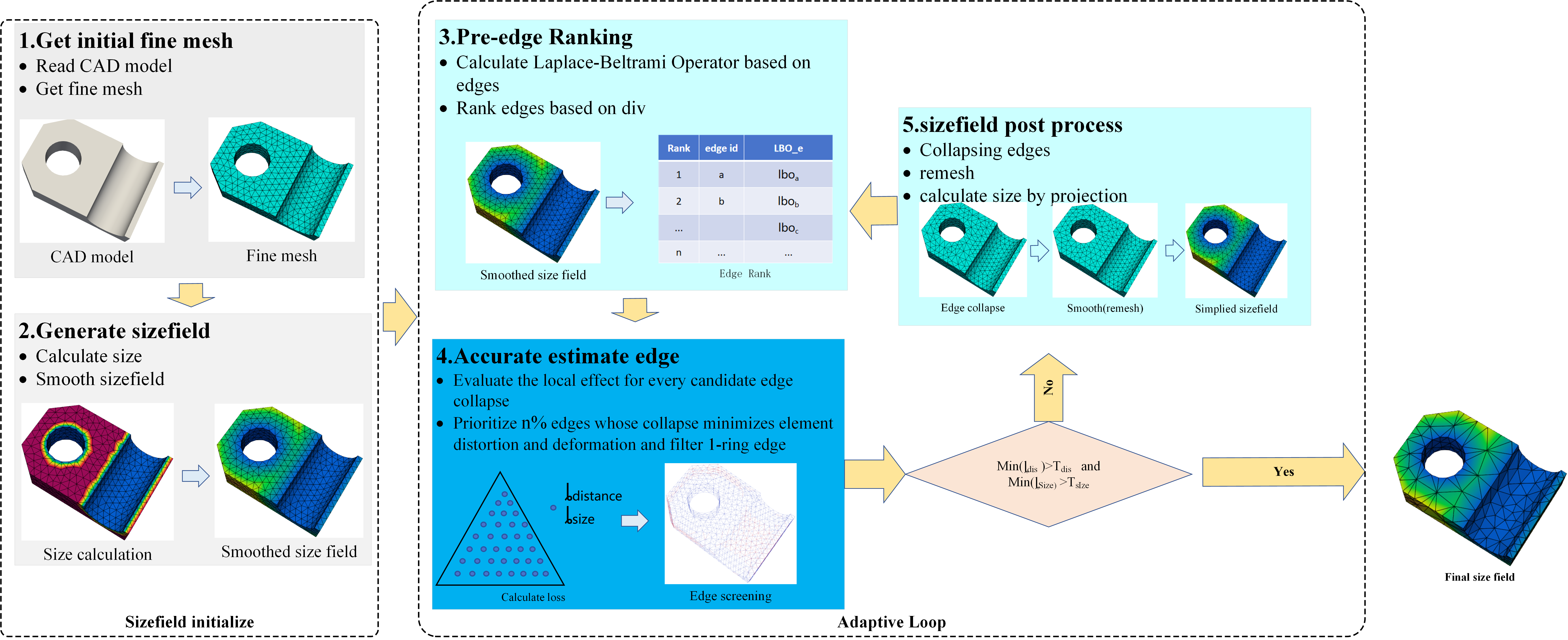}
		\caption{\centering{The workflow of the procedural adaptive background grid coarsening framework.}}
		\label{fig:adaptive1}
	\end{figure*}
	
	\subsubsection{Size Field Initialization}
	The core idea of this step is to generate an initial high-quality grid for subsequent iteration. First, a constant sizing field is created by setting a relatively small, uniform size value. This sizing field is then used to generate an initial background grid by invoking a surface meshing algorithm \cite{liuRobustFastLocal2025,yuEfficiencyAdvancingfrontSurface2022,yuMinimalSurfaceguidedHigherorder2024}. The primary purpose of this procedure is to produce a grid that is sufficiently geometry-preserving and possesses high element quality. This initial grid is intentionally dense and free from banding artifacts; should such artifacts appear, the uniform size value can be further decreased.
	
	Next, a sizing field is generated on this initial fine background grid. The process is generally divided into two steps, as illustrated in Figure \ref{fig:adaptive1}:
	\begin{enumerate}
		\item \textbf{Step 1:} An initial field $H_0$ is calculated based on geometric definitions such as curvature and proximity \cite{chenAutomaticSizingFunctions2017}.
		\item \textbf{Step 2:} The size field is smoothed subject to a gradient limit.
	\end{enumerate}
	The resulting field serves as an approximation of the ideal field $H^*$ and provides the guiding criterion for grid simplification in subsequent stages.
	
	This strategy of starting with a dense grid and then coarsening it offers significant advantages over generating grids directly from STL files, which often suffer from poor geometric conformity, banding, and low element quality. Our approach effectively circumvents these common defects from the outset. As a result, the optimization objective for the coarsening process is simplified: the goal is to minimize the total number of elements while maintaining the initial high-quality, defect-free state.
	
	\subsubsection{Pre-edge Ranking}
\begin{figure}[h]
		\centering
		\begin{subfigure}[b]{0.5\textwidth}
			\centering
			\includegraphics[width=0.75\textwidth]{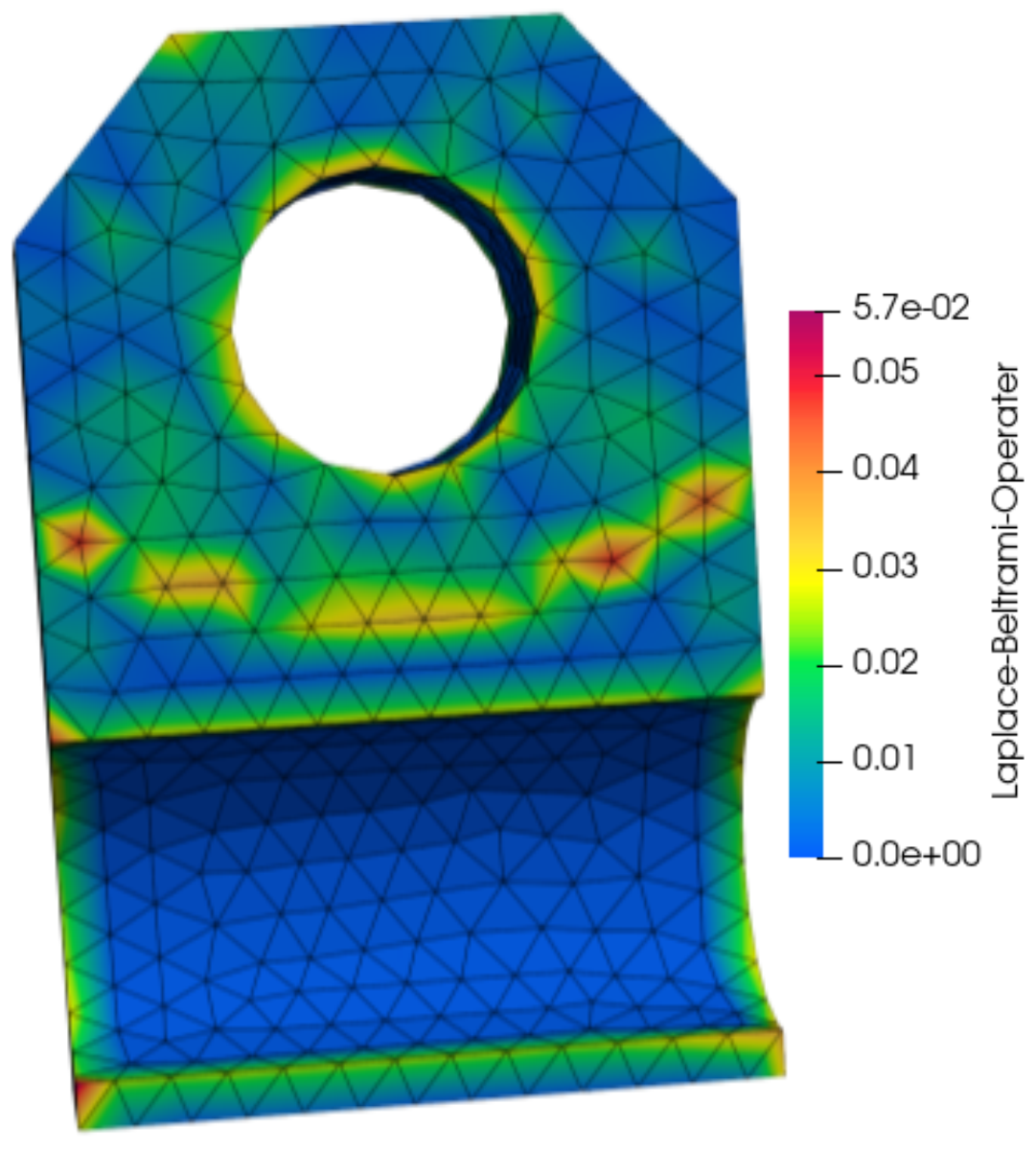}
			\caption{The Laplace-Beltrami operator on a vertex}
			\label{fig:dl-bo-sub1}
		\end{subfigure}
		\begin{subfigure}[b]{0.4\textwidth}
			\centering
			\includegraphics[width=\textwidth]{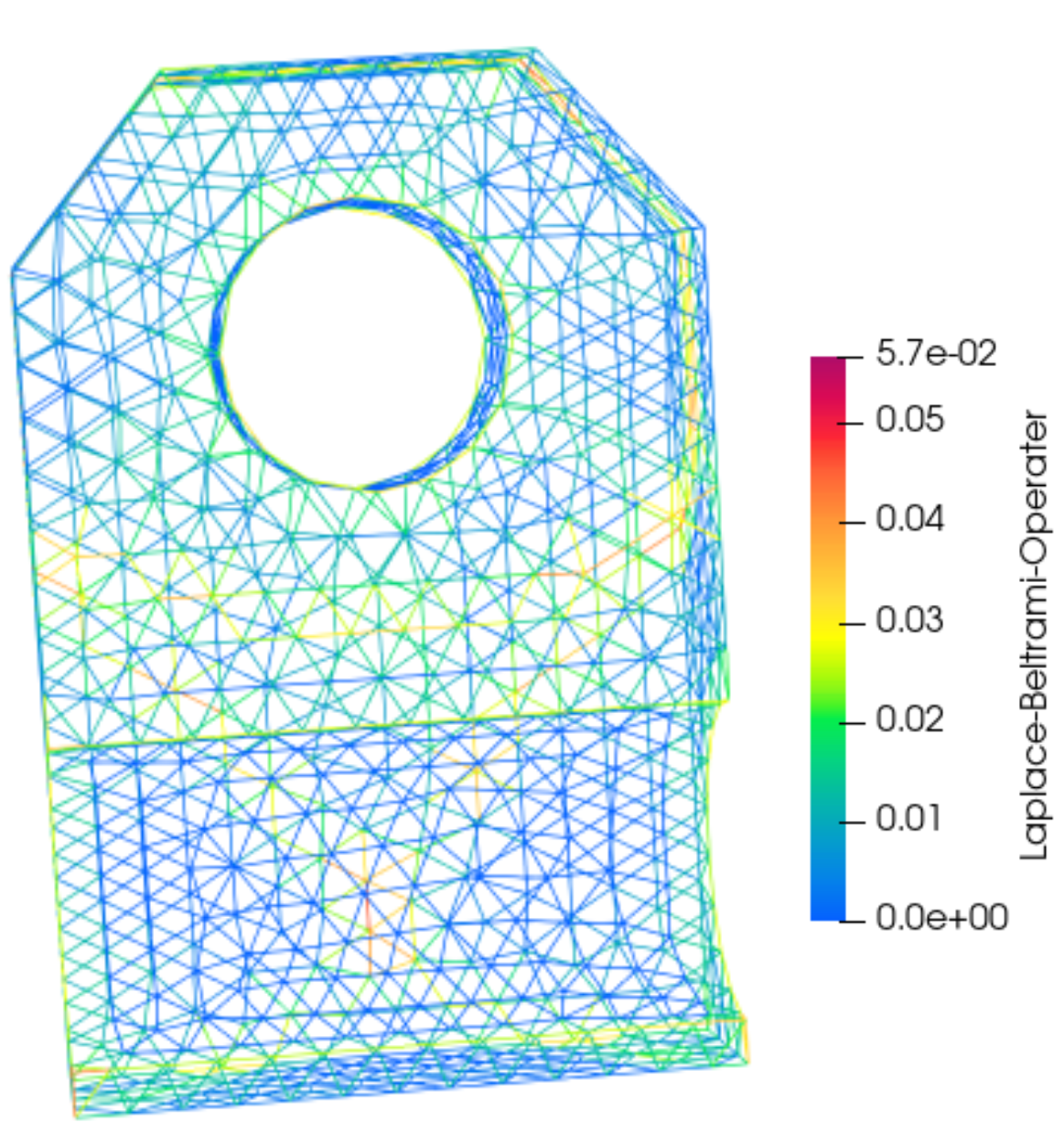}
			\caption{The Laplace-Beltrami operator on an edge}
			\label{fig:dl-bo-sub2}
		\end{subfigure}
		\caption{The gradient of the sizing field.}
		\label{fig:dl-bo}
	\end{figure}

	The \textbf{Pre-edge Ranking} stage serves as an efficient mechanism to prioritize all edges for potential simplification. Instead of computationally expensive Hessian matrix calculations, our approach ranks edges based on the local smoothness of the size field. This allows the subsequent, more accurate evaluation to focus on the most promising candidates. As illustrated in the flowchart, this process involves two main steps:
	\begin{enumerate}
		\item For each edge of the mesh defined on the smoothed size field, we compute a metric based on the discrete Laplace-Beltrami Operator (LBO), as illustrated in Figure \ref{fig:dl-bo}. 
		This metric, denoted as $LBO_e$, is defined as the average of the LBO values at the two vertices of the edge (Figure \ref{fig:dl-bo-sub1}). 
		The value of $LBO_e$ serves as an approximation of the local divergence of the size field along that edge, a concept depicted in Figure \ref{fig:dl-bo-sub2}.
		\item Next, all edges are sorted and ranked based on this divergence value ($LBO_e$) in ascending order. This procedure generates a complete, prioritized list, as depicted in the \textbf{"Edge Rank" table}. Edges in the smoothest regions (with the smallest divergence) receive the highest rank.
	\end{enumerate}
	
	The output of this stage is a \textbf{complete ranked list} of all edges, which is then passed as input to the "Accurate Edge Estimation" stage for a more rigorous and localized evaluation.
	
	\subsubsection{Accurate Edge Estimation}\label{sec:accurate_est}
	The \textbf{Accurate Edge Estimation} stage performs a rigorous evaluation of candidate edges to select an optimal set for simplification. For every candidate edge identified in the previous step, the algorithm assesses the local effect of a hypothetical collapse. This evaluation involves calculating a composite "loss" based on two key metrics, as illustrated in the "Calculate loss" diagram:
	\begin{enumerate}
		\item \textbf{Geometric Distortion ($l_{\text{distance}}$):} This metric quantifies the geometric error. It is computed by sampling points within the local triangular patch of the edge and measuring their projection distance to the original, high-fidelity grid after the virtual collapse.
		\item \textbf{Size Field Deviation ($l_{\text{size}}$):} This metric measures the deviation from the target size field. It is calculated by comparing the size values at the same sample points on the post-collapse grid with those on the ground-truth size field.
	\end{enumerate}
	
	The calculation of $l_{\text{distance}}$ and $l_{\text{size}}$ for each edge is presented in Algorithm \ref{alg:labeling}. For each edge $e$, a hypothetical post-collapse grid, $M_{\text{data}}$, is generated. Points are then uniformly sampled on the triangular faces adjacent to the edge, avoiding the triangle boundaries. These sample points are projected onto both the high-density target background grid, $M_{\text{target}}$, and the post-collapse grid, $M_{\text{data}}$, to compute the target size $s_{\text{target}}$, the current size $s_{\text{data}}$, and the projection distance $d_i$ for each point. Subsequently, the size variation, $\Delta_s$, is calculated based on the maximum ratio between $s_{\text{target}}$ and $s_{\text{data}}$ across all sample points. The distance variation, $\Delta_d$, is the maximum of the projection distances $d_i$.
	
	Based on this combined loss, the algorithm then \textbf{prioritizes the top $n$\% of edges} whose collapse would introduce the minimum combined element distortion and size field deformation. If $\Delta_s < T_{size}$ and $\Delta_d < T_{dis}$, the edge is considered collapsible. Finally, to prevent the formation of degenerate elements from simultaneous collapses in the same region, a \textbf{1-ring filter} is applied to this prioritized list, ensuring no two selected edges are adjacent. The output, shown as "Edge screening," is a refined, spatially distributed set of edges ready for the collapse operation. If no edge is selected, the adaptive loop is terminated.
	
	\begin{algorithm}[htbp]
		\caption{Edge Collapsing Loss Calculation}
		\label{alg:labeling}
		\begin{algorithmic}[1]
			\REQUIRE
			Labeled grid $M_{\text{label}}$, target grid $M_{\text{target}}$
			\ENSURE
			Size loss $\mathbf{s}$, distance loss $\mathbf{d}$
			\STATE Initialize loss vectors $\mathbf{s} \gets \emptyset$, $\mathbf{d} \gets \emptyset$
			\FOR{Every edge $e_i \in E(M_{\text{label}})$}
			\STATE $M_{\text{data}} \gets \text{EdgeCollapse}(M_{\text{label}}, e_i)$
			\STATE $T \gets \text{GetAdjacentFaces}(e_i, M_{\text{data}})$
			\STATE $P \gets \text{SamplingPoints}(T)$
			\STATE $s_{\text{data}} \gets \text{EstimateLocalSize}(P, M_{\text{data}})$
			\STATE $s_{\text{target}} \gets \text{EstimateLocalSize}(P, M_{\text{target}})$
			\STATE $d_i \gets \text{DistanceToSurface}(P, M_{\text{target}})$
			\STATE $\mathbf{s}[i] \gets \max(\frac{s_{\text{data}}}{s_{\text{target}}}, \frac{s_{\text{target}}}{s_{\text{data}}})$
			\STATE $\mathbf{d}[i] \gets \max(d_i)$
			\ENDFOR
			\RETURN $\mathbf{s}, \mathbf{d}$
		\end{algorithmic}
	\end{algorithm}
	
	\subsubsection{Size Field Post-processing}
	The \textbf{Size Field Post-processing} stage is the executive phase of the adaptive coarsening loop, where the background grid is actively simplified and its quality is restored. This stage is performed after the optimal candidate edges for collapse have been identified. The process comprises three sequential operations:
	\begin{enumerate}
		\item \textbf{Edge Collapsing:} The procedure begins by executing the collapse operations on the prioritized edges. As depicted in the "Edge collapse" sub-figure, this action merges the two vertices of an edge into a single new vertex, thereby reducing the total number of vertices and faces in the grid. This is the primary mechanism for coarsening the size field representation.
		\item \textbf{Remeshing and Smoothing:} Following the topological changes from edge collapses, which can introduce poorly shaped elements, a remeshing and smoothing step is applied. This procedure, illustrated in the "Smooth(remesh)" sub-figure, adjusts vertex positions and may perform local topological optimizations (such as edge-flipping) to improve the overall quality of the grid elements. The goal is to restore a well-conditioned grid with elements that are as close to equilateral as possible.
		\item \textbf{Size Recalculation by Projection:} After the grid topology and vertex positions have been updated, the size values at the new and moved vertices must be redefined. This is accomplished by projecting the vertices of the newly coarsened grid back onto the original, high-density fine grid. The size value at each vertex is then interpolated from the corresponding location on the original grid. This ensures that the simplified size field continues to accurately represent the geometric and physical requirements of the original model.
	\end{enumerate}
	Upon completion, the resulting simplified size field is fed back into the "Pre-edge Ranking" stage for the next iteration of the adaptive loop, unless a termination criterion is met.
	
	\subsection{GCN-Accelerated Framework for Adaptive Simplification}\label{sec:gnn}
	\begin{figure*}[h]
		\centering
		\includegraphics[width=\textwidth]{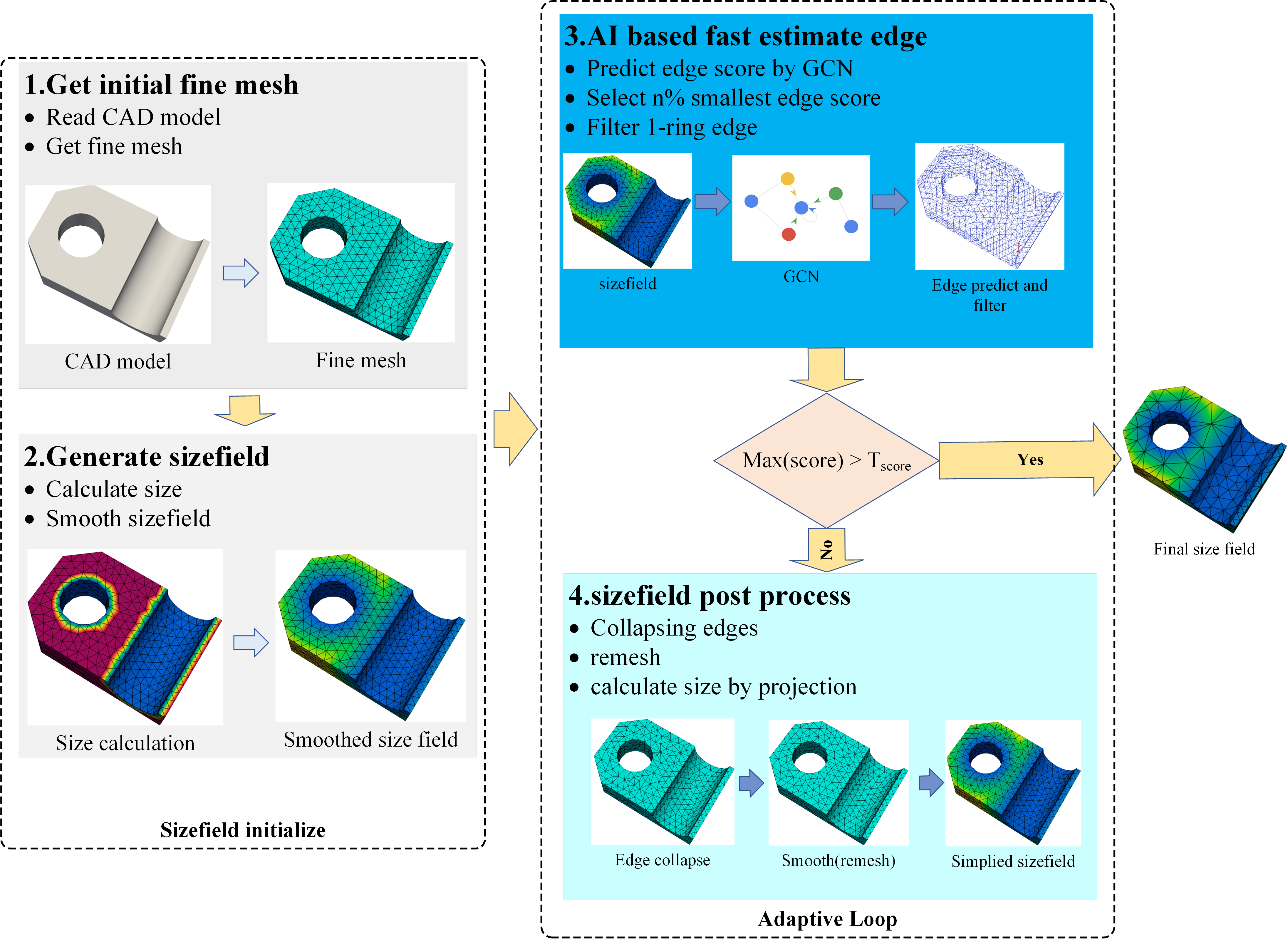}
		\caption{\centering{The workflow of the GCN-based adaptive background grid coarsening framework.}}
		\label{fig:adaptive2}
	\end{figure*}
	
	\subsubsection{Application of GCN}
	The procedural estimation of edge collapse effects, as discussed in Section \ref{sec:accurate_est}, is robust but introduces a significant computational bottleneck, making it prohibitively time-consuming for large-scale grids. To overcome this limitation, we introduce a novel, AI-driven methodology for rapidly identifying optimal edges for simplification. As illustrated in Figure \ref{fig:adaptive2}, we replace the conventional evaluation with a stage termed "AI-based fast edge estimation," which is powered by a GCN.
	
	The core of this approach is to reframe the edge selection problem from one of explicit calculation to one of learned inference. Instead of algorithmically computing a collapse cost for each edge, the GCN is trained to function as a highly efficient surrogate model. It learns the complex mapping from the local geometric and size field features of the grid directly to a predictive "edge score." This score serves as a proxy for the edge's suitability for collapse, where a lower score indicates a more favorable candidate.
	
	The application of the GCN within the adaptive loop follows a clear, three-step process:
	\begin{enumerate}
		\item \textbf{Predict Edge Score by GCN:} In a single forward pass, the trained GCN takes the current size field as input and holistically predicts a score for every edge in the grid simultaneously.
		\item \textbf{Select Candidates:} A predefined fraction (n\%) of the edges with the smallest scores are selected as the initial candidates for simplification.
		\item \textbf{Filter Conflicts:} A 1-ring filter is then applied to this candidate set to ensure that no two selected edges are immediately adjacent, thereby preventing localized grid degradation.
	\end{enumerate}
	The primary advantage of this GCN-based method is the profound improvement in computational efficiency. The costly, iterative, per-edge evaluation is replaced by a single, highly parallelizable neural network inference. This allows the adaptive loop to perform more iterations in less time, leading to much faster convergence towards the final size field. The process terminates when the score of the best candidate edge exceeds a predefined threshold, $T_{score}$.
	
	\subsubsection{Feature Vector}
\begin{figure*}[h]
		\centering
		\includegraphics[width=\textwidth]{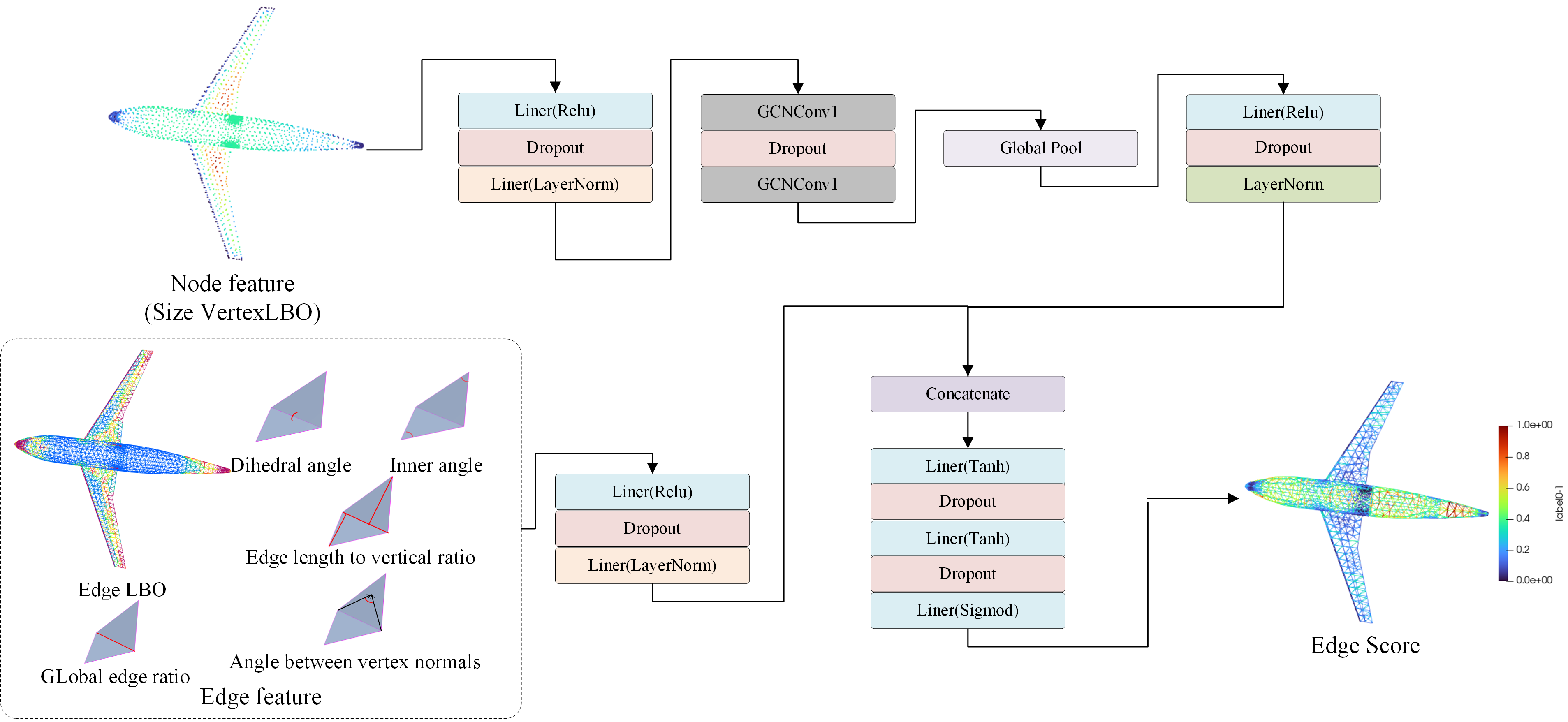}
		\caption{\centering{The dual-branch architecture of the Graph Convolutional Network.}}
		\label{fig:gcn}
	\end{figure*}

	As illustrated in Figure \ref{fig:gcn}, the model utilizes both nodes (vertices) and edges of the grid as input, defining distinct feature vectors for each to capture both global and local information.
	
	\paragraph{Node Features}
	Each node (vertex) is characterized by a two-dimensional feature vector comprising two key components:
	\begin{enumerate}
		\item \textbf{Size:} This value represents the desired element size in the vicinity of the node, providing the model with global information about the target mesh density.
		\item \textbf{Vertex LBO:} This is the value of the \textbf{Laplace-Beltrami Operator (LBO)} computed at each vertex, which reflects the local geometric smoothness or curvature.
	\end{enumerate}
	Hence, the node feature vector is represented as $[\text{Size}, \text{VertexLBO}]$.
	
	\paragraph{Edge Features}
	Each edge is described by an eight-dimensional feature vector that combines a smoothness metric with local geometric attributes:
	\begin{enumerate}
		\item \textbf{Edge LBO:} The value of the Laplace-Beltrami Operator computed on the edge, quantifying the smoothness of the size field in its local region.
		\item \textbf{Dihedral Angle:} The angle between the two triangular faces adjacent to the edge.
		\item \textbf{Inner Angles:} The two angles within the adjacent triangles that are opposite the edge.
		\item \textbf{Edge Length to Height Ratios:} The ratios of the edge's length to the heights of the opposing vertices in the two adjacent triangles, which reflects element quality.
		\item \textbf{Global Edge Ratio:} The ratio of the edge's length to the average length of all edges in the grid. This feature provides a global geometric context.
		\item \textbf{Angle Between Vertex Normals:} The angle between the surface normals at the two endpoint vertices of the edge, describing the local surface curvature.
	\end{enumerate}
	This multi-dimensional feature vector provides the model with a rich geometric and topological description of each edge and its immediate neighborhood.
	
	\subsubsection{Model Architecture}
	The GCN employs a \textbf{dual-branch architecture} to process node and edge features in parallel before fusing them for a final prediction.
	
	\paragraph{Branch 1: Node Feature Branch (Global Context Path)}
	This branch is designed to learn a global context representation from the node features.
	\begin{enumerate}
		\item \textbf{Initial Embedding:} The 2D node feature vectors are first passed through a Multi-Layer Perceptron (MLP) block, consisting of a \texttt{Linear(ReLU)} layer, a \texttt{Dropout} layer, and a \texttt{Linear(LayerNorm)} layer. This block embeds the input features into a higher-dimensional latent space.
		\item \textbf{Graph Convolution:} The resulting node embeddings are processed by two consecutive \texttt{GCNConv} layers. These layers aggregate information from neighboring nodes, enabling the network to learn local structural patterns and propagate information across the grid.
		\item \textbf{Global Pooling:} A \texttt{Global Pool} layer then aggregates all node feature vectors into a single, fixed-size global context vector, which serves as a condensed summary of the entire grid's properties.
		\item \textbf{Global Feature Refinement:} This global context vector is further processed by another MLP block for additional non-linear transformation and refinement.
	\end{enumerate}
	
	\paragraph{Branch 2: Edge Feature Branch (Local Feature Path)}
	This branch independently processes the local geometric features of each edge.
	\begin{itemize}
		\item \textbf{Edge Feature Embedding:} The 8D edge feature vectors are fed into an MLP block, similar in structure to the initial embedding layer of the node branch, to generate a high-dimensional local feature embedding for each edge.
	\end{itemize}
	
	\paragraph{Feature Fusion and Prediction Head}
	To ensure that the prediction for each edge is informed by both local geometry and global context, the information from the two branches is fused.
	\begin{enumerate}
		\item \textbf{Concatenation:} The global context vector from the node branch is \textbf{concatenated} with the local feature embedding of \textit{each} edge from the edge branch. This creates a unified feature vector for every edge that contains both local and global information.
		\item \textbf{Prediction Head:} The concatenated feature vectors are passed through a final MLP prediction head. This head culminates in a final \texttt{Linear} layer with a \textbf{Sigmoid} activation function. The Sigmoid function maps the output to a normalized \textbf{Edge Score} within the range $[0, 1]$, representing the predicted suitability of each edge for collapse.
	\end{enumerate}
	
	\section{Dataset}
	All geometric model data used in this study are derived from publicly available, high-quality open-source datasets to ensure reproducibility. Component models with relatively simple structures are selected from the STEP-format CAD models provided by the ABC dataset (https://deep-geometry.github.io/abc-dataset/). This dataset is a widely recognized benchmark in computer graphics and geometric processing.
	
	To evaluate the algorithm's performance on complex assemblies, we adopt circuit models from the GrabCAD open-source community (https://grabcad.com). The selected circuit models consist of multiple parts, intricate structures, and assembly relationships, enabling an effective assessment of the algorithm’s adaptability in complex scenarios.
	
	As shown in Figures \ref{fig:adaptive1} and \ref{fig:adaptive2}, the procedure begins with the discretization of the CAD model into a surface mesh, upon which the sizing field is then calculated. For the initial fine grid generation and size value calculation, we employ the TiGER mesh generation engine, which incorporates a surface meshing algorithm based on the Advancing Front Technique (AFT) as well as a lightweight geometric engine.
	
	By integrating structured models from standardized datasets with complex assembly models from engineering practice, we construct a test set that is both representative and diverse, thereby providing a solid foundation for a comprehensive evaluation of the proposed method. All of the code and dataset can be found in https://github.com/zhuxy42-oss/ABGS.git.
	
	\section{Experiments}
	To validate the efficacy of the proposed framework, we have designed a series of experiments on several representative test cases. The evaluation is structured into three main parts. First, we assess the predictive efficacy of the GCN in the edge scoring task. Second, we evaluate the overall performance of the integrated simplification framework using key metrics: the number of elements in the background grid (query elements), the element count of the final surface mesh, and the degree of geometric fidelity preservation. Finally, we demonstrate the practical utility of our GCN-based framework by analyzing its computational efficiency compared to the procedural method.
	
	\subsection{Experiment Setup}
	All models were implemented in PyTorch and trained for 1200 epochs on a single NVIDIA RTX 3090 GPU. We utilized the Adam optimizer with an initial learning rate of $10^{-4}$, managed by a stepwise learning rate scheduler. The network architecture consists of a node encoder and an edge encoder (each with two 128-neuron hidden layers), followed by two graph convolutional layers (128 neurons each). A context encoder (one 128-neuron hidden layer) then aggregates features for a final edge scorer, which is a multilayer perceptron with three hidden layers of dimensions [512, 256, 128]. The sigmoid activation function was used in all layers. We denote the algorithms from Section \ref{sec:procedural} and Section \ref{sec:gnn} as LBO-ABGS and GCN-ABGS, respectively, to facilitate discussion and citation.
	
	\subsection{Validation of Stability and Continuity in the Adaptive Framework}
	To validate the stability and continuity of our adaptive framework, this section analyzes the coarsening process by monitoring the \textbf{total number of mesh elements} and the \textbf{number of query cells in the background grid}. The number of query cells is critical, as insufficient density can impair the capture of local details, leading to "banding artifacts" and potential numerical instability.
	
\begin{figure*}[htbp]
		\centering
		\includegraphics[width=\textwidth]{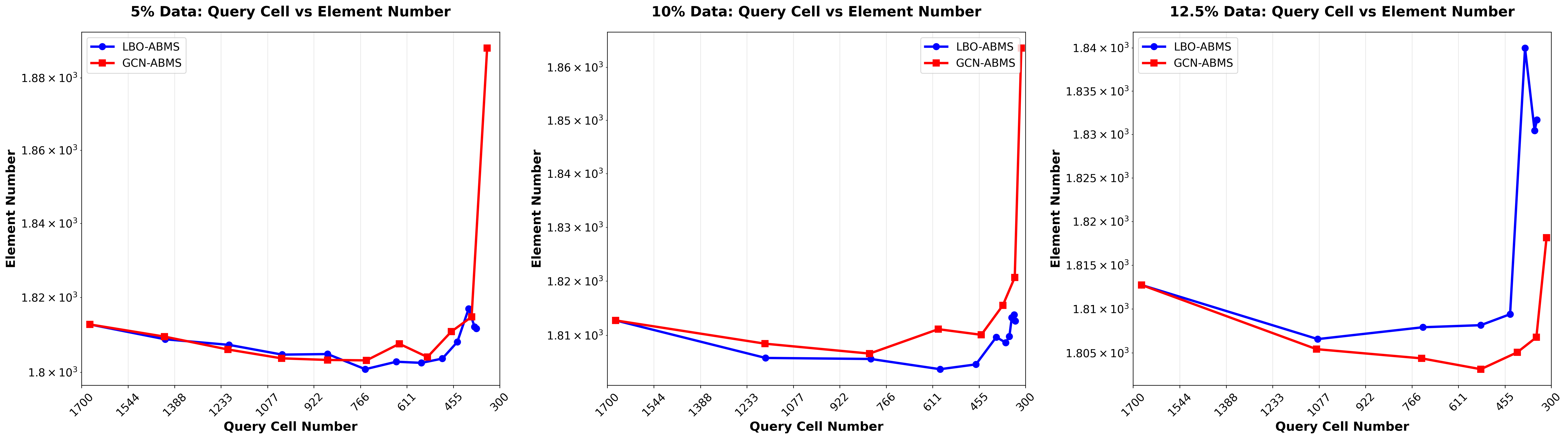}
		\caption{\centering{Adaptive coarsening process for LBO-ABGS (left) and GCN-ABGS (right). The y-axis shows the total number of surface mesh elements, and the x-axis shows the number of background grid query cells.}}
		\label{fig:adaptive}
	\end{figure*}

	Figure \ref{fig:adaptive} illustrates the performance difference between LBO-ABGS and GCN-ABGS. In these plots, the x-axis represents the number of query cells in the background grid during coarsening, while the y-axis indicates the total number of elements in the corresponding surface mesh. The three curves in each plot correspond to the convergence behavior when simplifying 5\%, 10\%, and 12.5\% of the edges per iteration, respectively.
	
	Both plots show that the total element count remains relatively stable during the initial coarsening phase. However, once the number of query cells falls below a critical threshold (approximately 455 in this case), all curves exhibit a significant divergent trend. This occurs because an overly sparse background grid fails to accurately represent the geometry, resulting in severe banding.
	
	Comparing the two methods, GCN-ABGS process exhibits a faster rate of divergence in the late stages of coarsening. This can be attributed to the inherent uncertainty in the GCN's predicted scores. Since 100\% prediction accuracy is not guaranteed, prediction errors can accumulate over many iterations, exacerbating the exponential growth of the element count when the background grid becomes severely deficient.
	
	Nevertheless, this divergent behavior only occurs when the background grid is extremely sparse—a scenario averted in practice by implementing a suitable termination criterion. The overall evolution curves demonstrate that our framework maintains high stability before this divergent phase. Therefore, by setting an appropriate stopping condition, we can generate a high-quality background grid that meets requirements while ensuring stability.
	
	\subsection{Performance of Background Grid Simplification}
\begin{figure*}[htbp]
		\centering
		\includegraphics[width=\textwidth]{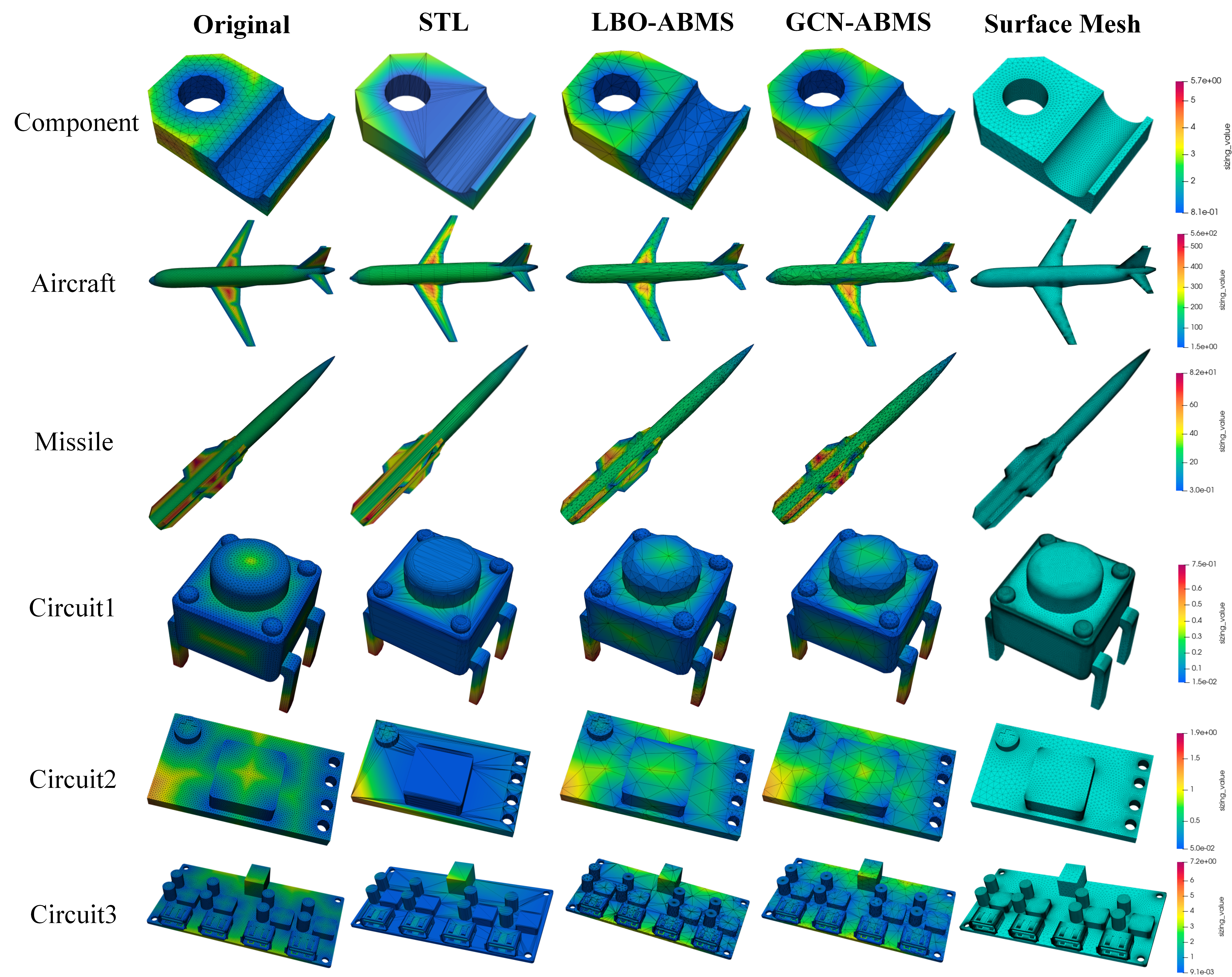}
		\caption{\centering{Visual comparison of background grid simplification results across various models.}}
		\label{fig:bkgm-simplification}
	\end{figure*}

	This section evaluates the effectiveness of the proposed simplification method using three metrics: the number of generated surface mesh elements, the number of queryable background grid elements, and the Hausdorff distance to the original dense grid. Table \ref{tab:experiment} and Figure \ref{fig:bkgm-simplification} present the results.
	
	The proposed method achieves a reduction of 74.48\%–93.73\% in the number of background grid elements across multiple test models. This substantial reduction greatly decreases computational resource consumption.
	
	To assess geometric fidelity, we use the Hausdorff distance to measure the deviation between the simplified and the original dense background grids. As shown in Figure \ref{fig:bkgm-simplification}, the simplified grids exhibit negligible perceptual differences, and the Hausdorff distances reported in Table \ref{tab:experiment} remain within acceptable bounds. This demonstrates that our method achieves high-efficiency simplification while maintaining excellent geometric conformity.
	
	The ultimate purpose of the background grid is to guide high-quality surface mesh generation. The number of surface mesh elements generated using simplified backgrounds deviates by no more than 17\% from the results obtained using the original dense backgrounds. This indicates that the simplified sizing fields remain stable and reliable for generating high-quality surface meshes.
	
	In comparison with STL-based background grids, our simplified grids are free from the severe banding phenomena that often plague STL files. While STL grids may have fewer elements, banding leads to an over-refinement of the final surface mesh in flat regions. Our method avoids this, producing more efficient surface meshes.
	
	A comparison between LBO-ABGS and GCN-ABGS methods reveals that the LBO-ABGS approach often achieves slightly more aggressive simplification. This is because it explicitly computes the cost of each collapse, whereas the GCN predicts scores based on learned weights, which can introduce minor errors. However, the difference in the number of resulting surface mesh elements is small (less than 8\%), making the GCN a highly effective and practical alternative.
	
	\begin{table}[t]
		\centering
		\scriptsize 
		\begin{tabular}{lccccccc}
			\hline
			Metric & Condition & Component & Aircraft & Missile & Circuit1 & Circuit2 & Circuit3\\
			\hline
			\multirow{4}{*}{Surface Mesh cell} 
			& Begin & 8758 & 197766 & 272570 & 239850 & 44820 & 629057\\
			& LBO-ABGS & 8658 & 189576 & 317694 & 180218 & 41444 & 657510\\
			& GCN-ABGS & 8598 & 204712 & 327504 & 189464 & 42122 & 657307\\
			& STL & 10910 & 268700 & 415480 & 380544 & 172040 & 2123802\\
			\hline
			\multirow{4}{*}{background grid cell} 
			& Begin & 1672 & 19230 & 37716 & 12242 & 26154 & 65617\\
			& LBO-ABGS & 454 & 2102 & 3945 & 1666 & 1666 & 10085\\
			& GCN-Adaptive & 594 & 2112 & 8948 & 1895 & 2058 & 12457\\
			& STL & 200 & 2849 & 2286 & 3958 & 1214 & 11974\\
			\hline
			\multirow{4}{*}{Hausdorff} 
			& Begin & 2.34 & 148.59 & 9.59 & 0.03 & 0.07 & 0.33\\
			& LBO-ABGS & 6.50 & 160.59 & 13.24 & 0.16 & 0.07 & 0.33\\
			& GCN-ABGS & 7.97 & 218.32 & 49.86 & 0.13 & 0.07 & 0.33\\
			& STL & 19.70 & 143.22 & 9.59 & 0.04 & 0.06 & 0.26\\
			\hline
		\end{tabular}
		\caption{Comparison of simplification results across different methods. "Begin" refers to the initial dense grid, and "STL" to using the raw STL file as the background grid.}
		\label{tab:experiment}
	\end{table}
	
	\subsection{Computational Performance}
	This section analyzes the computational efficiency of the entire workflow, from size-field simplification and smoothing to final volume mesh generation. Table \ref{tab:time-experiment} records the time consumption at each stage.
	
	As shown in the table, the simplified background grids exhibit significant improvements in query speed. For complex models like aircraft and circuits, query times decrease by 56\%–88\%. This is due to the drastic reduction in element count, which lowers the complexity of spatial searches.
	
	The time required for gradient smoothing is also significantly reduced. For grids simplified by LBO-ABGS, smoothing time is reduced by 71\%–94\% compared to the original grids. This is attributed to the smaller number of vertices and faces, which markedly decreases the computational cost of the underlying convex optimization.
	
	A direct comparison of the two simplification algorithms reveals the dramatic efficiency advantage of GCN-ABGS. For all models, GCN-ABGS reduces the simplification runtime by one to two orders of magnitude compared with LBO-ABGS. This speed advantage stems from the GCN's learning-driven predictive paradigm: after offline training, the model infers a simplification strategy in a single efficient forward pass, whereas LBO-ABGS requires costly iterative computation for each edge. This order-of-magnitude speedup establishes the unmatched practicality and scalability of the GCN-ABGS for large-scale grid problems.
	
	Compared to LBO-ABGS and GCN-ABGS, the STL-based method exhibits a faster query time due to its coarser grid structure with fewer cells. However, it requires substantially more time for both smoothing and final mesh generation. The prolonged smoothing process is caused by the presence of numerous poor-quality elements, particularly those with large angles. Furthermore, the extended generation time is attributable to geometric artifacts, such as banding elements, which increase the computational complexity for the mesh generator.
	
	\begin{table}[t]
		\centering
		\scriptsize 
		\begin{tabular}{lccccc}
			\hline
			Metric & Time (s) & Component & Aircraft & Missile & Circuit1 \\
			\hline
			\multirow{4}{*}{Begin} 
			& Smooth & 26.30 & 1291.63 & 6070.30 & 8015.25 \\
			& Query & 17.64 & 104.98 & 17.75 & 23.06  \\
			& Generate & 194.37 & 754.39 & 163.69 & 71.31 \\
			& Total & 220.67 & 2046.02 & 6233.99 & 8086.56\\
			\hline
			\multirow{5}{*}{LBO-ABGS} 
			& Smooth & 1.49 & 165.11 & 1703.02 & 100.27\\
			& Simplify & 118.50 & 6775.84 & 19177.32 & 7880.71 \\
			& Query & 11.40 & 36.14 & 7.08 & 2.58\\
			& Generate & 152.48 & 844.60 & 153.22 & 41.29 \\
			& Total & 272.38 & 7821.69 & 21033.64 & 8024.85 \\
			\hline
			\multirow{5}{*}{GCN-ABGS} 
			& Smooth & 1.08 & 135.35 & 1535.40 & 117.82\\
			& Simplify & 13.47 & 105.67 & 188.03 & 96.04  \\
			& Query & 11.24 & 69.35 & 7.33 & 2.70 \\
			& Generate & 153.07 & 996.59 & 158.54 & 37.64 \\
			& Total & 178.86 & 1306.96 & 1889.30 & 254.20 \\
			\hline
			\multirow{4}{*}{STL} 
			& Smooth & 4.59 & 476.82 & 5178.11 & 435.85 \\
			& Query & 17.96 & 144.99 & 6.38 & 17.95  \\
			& Generate & 253.68 & 2079.15 & 185.2 & 157.30 \\
			& Total & 258.27 & 2555.97 & 5363.31 & 593.15\\
			\hline
		\end{tabular}
		\caption{Computational performance comparison (times in seconds). The "Total" for LBO-ABGS and GCN-ABGS includes the simplification time.}
		\label{tab:time-experiment}
	\end{table}
	
	\section{Conclusion}
	\label{sec:conclusion}
	
	This paper has presented a novel, GCN-based framework for the adaptive simplification of triangular background grids used in sizing field control. By reformulating the simplification task as a learned edge classification problem, our method replaces a computationally expensive procedural evaluation with a single, efficient forward pass of a GCN. This approach successfully addresses the key challenges of creating background grids that are computationally lightweight, geometrically conforming, and free from artifacts.
	
	Compared to conventional methods that rely on either raw STL triangulations or uniformly refined surface grids, GCN-ABGS achieves an excellent balance between computational performance and the quality of the final background grid. Experimental results on a diverse set of engineering models demonstrate significant reductions in background grid complexity (74\%-94\%) and subsequent sizing field query times (35\%-88\%), all while preserving the geometric fidelity required for generating high-quality final meshes. The substantial acceleration over LBO-ABGS highlights the practical value of our data-driven strategy for large-scale simulation workflows. Future work could explore the application of this framework to anisotropic sizing fields and its extension to volumetric background grids.
	
	\bibliographystyle{elsarticle-num-names} 
	\bibliography{my.bib}
	
	
	
	
	
	
\end{document}